\documentclass[useAMS,usenatbib]{mn2e}
\usepackage{graphicx}
\usepackage{epsfig}

\title[The Seyfert AGN RX J0136.9-3510 and the Spectral State of Super Eddington Accretion Flows]{The Seyfert AGN RX J0136.9-3510 and the Spectral State of Super Eddington Accretion Flows}

\author[C. C. Jin, C. Done, M. Ward, M. Gierli\'{n}ski, J. Mullaney]
{Chichuan Jin,
Chris Done, Martin Ward, Marek Gierli\'{n}ski,
James Mullaney\\
Department of Physics, University of Durham, South Road, Durham, DH1 3LE, UK}

\begin{document}

\maketitle

\label{firstpage}

\begin{abstract}

We have carried out a survey of long 50ks XMM-Newton observations of a
sample of bright, variable AGN. We found a distinctive energy
dependence of the variability in RXJ0136.9-3510 where the fractional
variability increases from 0.3 to 2~keV, and then remains
constant. This is in sharp contrast to other AGN where the X-ray
variability is either flat or falling with energy, sometimes with a
peak at $\sim$~2~keV superimposed on the overall trend. Intriguingly
these unusual characteristics of the variability are shared by one
other AGN, namely RE J1034+396, which is so far unique showing a
significant X-ray Quasi-Periodic Oscillation (QPO). In addition the
broad band spectrum of RXJ0136.9-3510 is also remarkably similar to
that of RE J1034+396, being dominated by a huge soft excess in the
EUV-soft X-ray bandpass. The bolometric luminosity of RX J0136.9-3510
gives an Eddington ratio of about 2.7 for a black hole mass (from the
H beta line width) of $7.9 \times 10^{7}M_{\sun}$. This mass is about
a factor of 50 higher than that of RE J1034+396, making any QPO
undetectable in this length of observation. Nonetheless, its X-ray
spectral and variability similarities suggest that RE J1034+396 is
simply the closest representative of a new class of AGN spectra,
representing the most extreme mass accretion rates.

\end{abstract}

\begin{keywords}
accretion, Eddington ratio, SED, active-galaxies: nuclei
\end{keywords}

\section{Introduction}

Active galactic nuclei (AGN) are powered by gas accreting onto the
central super-massive black hole. As such their intrinsic spectra are
determined simply by the mass and spin of the black hole, together
with the mass accretion rate, although their observational appearance
is also affected by absorption. The discovery that the black hole mass
scales with the host galaxy bulge properties (\citealt{Gebhardt00};
\citealt{Magorrian98}) presented a means to classify the bewildering
variety of unabsorbed AGN spectra in terms of two of these three
intrinsic properties. It was found that different optical spectral
types clearly correlated with black hole mass and mass accretion rate,
with the Narrow-line Seyfert 1 galaxies (NLS1s: defined as $H\beta$
emission line FWHMs less than 2000 km/s and $[O III]/H\beta$ flux
ratios less than 3.0 \citealt{Goodrich89}) as examples of lower mass
black holes accreting at higher Eddington ratio than the typical broad
line Seyfert 1s (BLS1s), which in turn had higher Eddington
ratio/lower black hole masses than the LINERSs
(e.g. \citealt{Osterbrock85}; \citealt{Boller96}; \citealt{Boroson02}; 
\citealt{Mathur00}; \citealt{Bian08}; \citealt{Winter09}).

NLS1's are intrinsically very luminous in the EUV/soft X-ray bandpass,
comprising over half the AGN found in the softest X-ray selected
surveys (\citealt{Puchnarewicz92}; \citealt{Grupe96};
\citealt{Edelson99}). This strong ionising flux results in intense
permitted optical Fe II emission lines from the broad line region and
high ionisation species often refered to as coronal lines
(\citealt{Wills99}; \citealt{Kuraszkiewicz00}; \citealt{Ghosh04};
\citealt{Mullaney08}). Other observed characteristics include rapid
X-ray variability (\citealt{Leighly99a}; \citealt{Boller96}) and
steeper soft X-ray spectra than those of BLS1s (\citealt{Leighly99b};
\citealt{Brandt97}). Taken together these distinctive properties have
led to them being considered as the scaled up counterparts of the most
luminous accretion states seen in stellar mass black hole binary (BHB)
systems ie. those having ``high'' and ``very high'' states
(\citealt{Pounds95}; \citealt{Middleton07}).

This link with the stellar mass black hole systems was strengthened by
the first detection of an X-ray QPO in the NLS1 RE J1034+396
(\citealt{Gierlinski08}) as QPO's are commonly seen in the BHB systems
(e.g. \citealt{Remillard06}). While there are multiple different types
of QPO in the BHB (high frequency QPO, low frequency QPO, plus some
others seen only in GRS1915+105: e.g. \citealt{Remillard06};
\citealt{Morgan97}), they are all generally most prominent in the
'very high state'. Simple scaling from BHB then predicts that QPO's
should be common amongst NLS1 as a class due to their high mass
accretion rate. Nontheless, RE J1034+396 is extreme even amongst NLS1
(see below), and it may be that its so far unique QPO detection is due
more to its unusual properties than to any simple scaling from the
known QPO's in BHB.

We first summarise the properties of RE~J1034+396, in order to provide
the context for our study of RX J0136.9-3510, which is reported in
this paper.  The most obvious unusual feature of RE~J1034+396 is its
broad band spectral energy distribution (SED). This exhibits a peak in
the far UV which connects smoothly onto the steep soft X-ray
spectrum. These components form a huge ``soft X-ray excess'' with
respect to the $\Gamma\sim 2.2$ X-ray tail which dominates above $\sim
2$~keV (\citealt{Puchnarewicz95}; \citealt{Casebeer06};
\citealt{Middleton07}; \citealt{Middleton09}).  The energy dependence
of this variability is also very different to that commonly seen in
other AGN. The fractional variability amplitude (as measured by root
mean square, hereafter {\it rms\/}) rises steeply to $\sim 2$~keV and
then levels off. This is most likely due to the presence of two
separate components in the X-ray spectrum, with the variability being
associated with the X-ray tail, whilst the soft excess component
remains more or less constant (\citealt{Middleton09}). This situation
contrasts with the flat or falling {\it rms\/} spectra seen in other
AGN, sometimes with a peak at $\sim 2$~keV superimposed on this
(\citealt{Vaughan03}; \citealt{Vaughan04}; \citealt{Fabian04};
\citealt{Gierlinski06}; \citealt{Ponti06}; \citealt{Petrucci07};
\citealt{Larsson08}), which makes it more likely that the apparent
soft excess in these objects is due instead to a single spectral
component distorted by reflection and/or absorption
(\citealt{Crummy06}; \citealt{Gierlinski06}).

In this paper we make use of the unusual energy dependence of the X-ray variability in
RE J1034+396 to search for potentially similar objects. A survey of
all long ($\ga 50$~ks) XMM-Newton observations of bright and variable AGN
yielded a similar {\it rms\/} shape only in one object RX J0136.9-3510 (2MASSi
J0136544-350952). This AGN also has a similar broadband spectrum
to RE~J1034+396, suggesting that they may form a subclass of the highest mass
accretion rate AGN.

\section{Source Selection}
We searched the {\it XMM-Newton Master Log\/} \& {\it Public
Archive\/} for pointed observations with exposure times in the PN
instrument of $\ge 50000$ s in ``subject{\_}category'' ''AGN, QSOs,
BL-Lacs and XRB''. This resulted in 115 observations available at the
time of our study (November 2008). We further refined
this criteria to include only those objects with PN count rates of
${\ga}~1.0$ counts/s, to include only bright sources for which
the variability can be well determined.

\begin{figure}
\epsfig{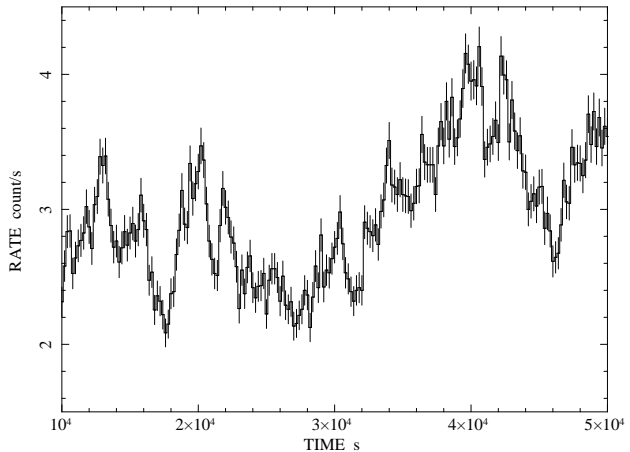}
\caption{Background subtracted lightcurve of RX J0136.9-3510 binned on 200 s. The exposure start time (UTC) is 2005-12-14 20:45:30, but the first 10 ks was excluded due to the high background contamination.}
\label{lightcurve}
\end{figure}

\begin{figure}
\includegraphics[scale=0.5,width=84mm]{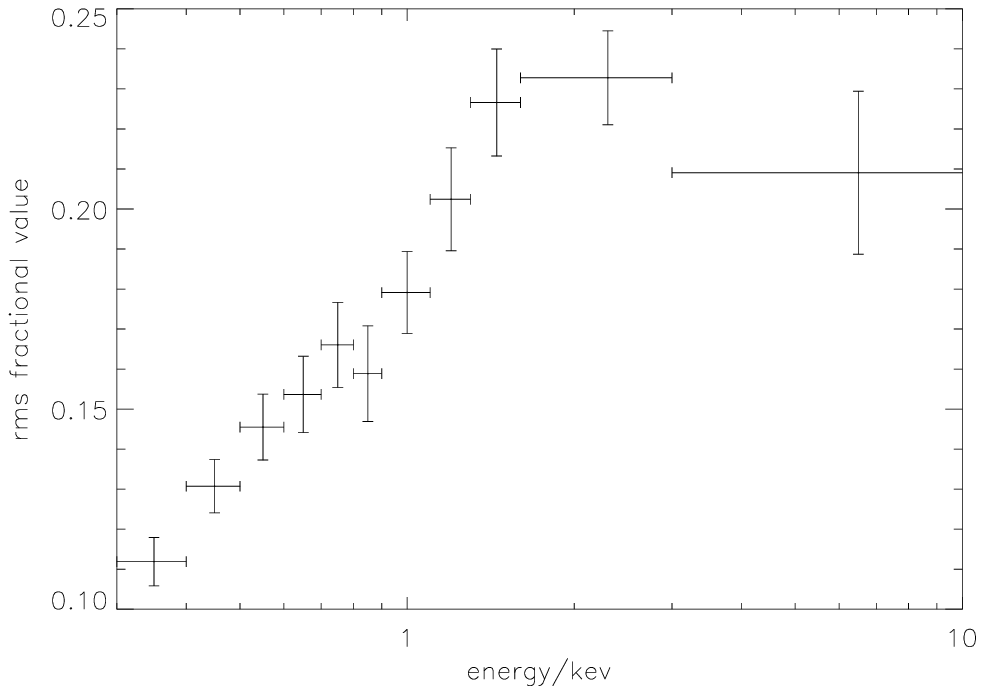}
\caption{The {\it rms\/} spectrum correspongding to Figure~\ref{lightcurve}, the binning time is 2000s.}
\label{rms-spectrum}
\end{figure}

\begin{figure}
\epsfig{file=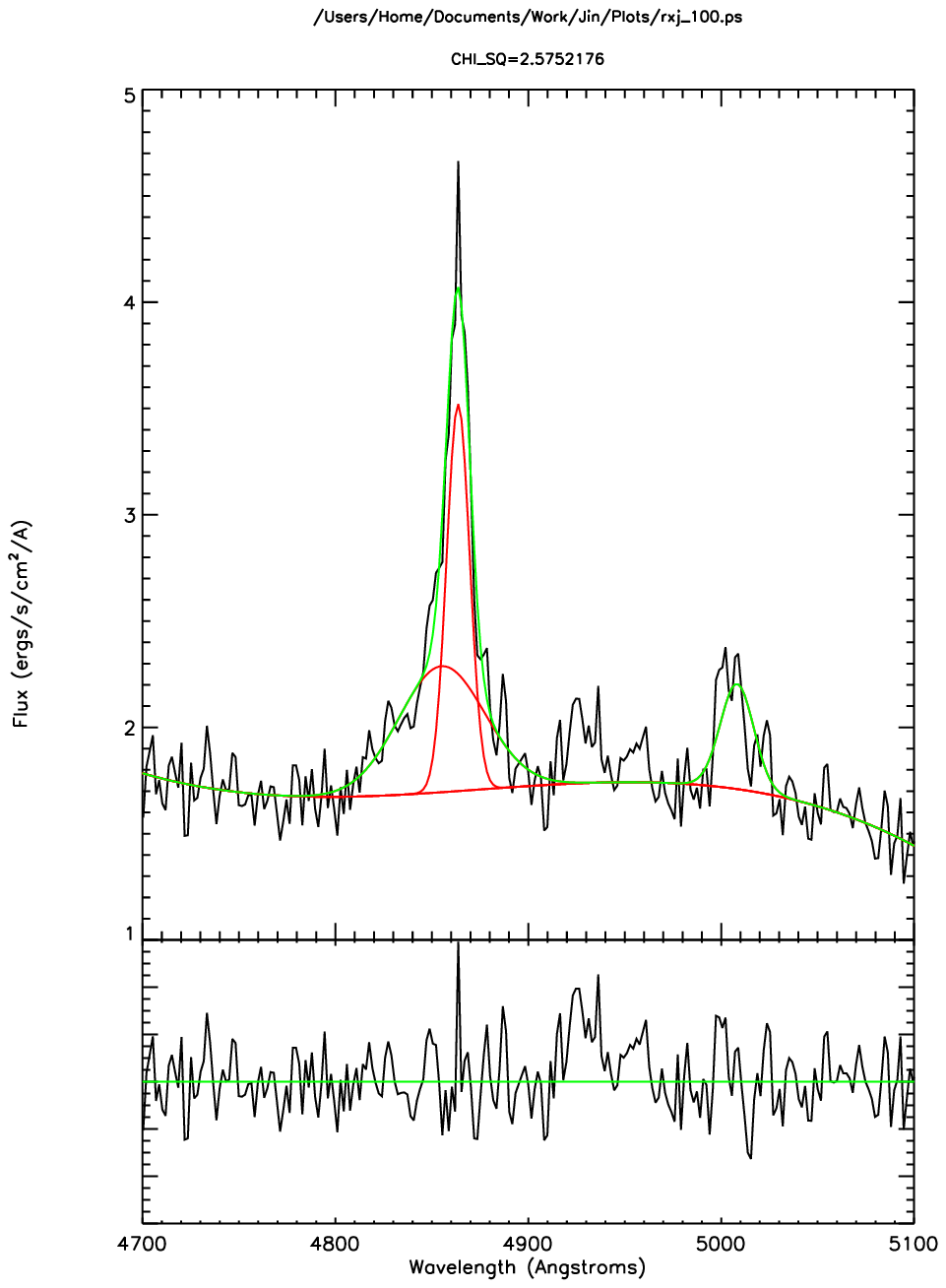, width=84mm, clip=}
\caption{The optical specturm of RX J0136.9-3510, including the H$\beta$ emission line fitting. The data is from Grupe D. and we show this figure with his permission.}
\label{optical-SED}
\end{figure}

\begin{figure}
\includegraphics[scale=0.5,width=84mm]{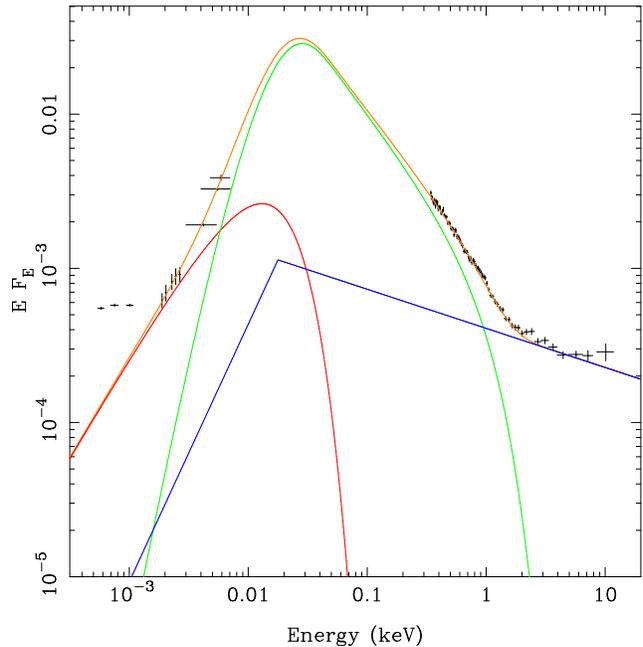}
\caption{RX J0136.9-3510 unfolded spectrum. All data points from different wave bands are included in this figure, though the infrared points are not included in the model fitting. The model spectrum without galactic extinction and dust reddening is also generated and superposed on the source spectrum, with red representing redshifted disk component, green representing {\it compTT\/} and blue representing {\it bknpower\/}. The orange line shows the total model spectrum.}
\label{SED}
\end{figure}

We also searched the {\it XMM-Newton Serendipitous Source Catalogue
(2XMMi Version)\/} for serendipious bright AGN detected in similarly
long exposures, by setting pn${\_}$8${\_}$flux ${\geqslant}10^{-12}$
erg cm$^{-2}$ s$^{-1}$ and pn${\_}$ontime${\geqslant}$50000 s. This
yielded 29 observations.

We then combined these two samples to give a
total of 68 individual sources. For each dataset we calculated the fractional variability amplitude
using {\tt lcstats} in {\tt xronos 5.21}.  This is defined as the root
mean square of the intrinsic (corrected for error bars) variance about
the lightcurve mean, $\sigma^2$, normalised to this mean, $I$. Only
sources which are strongly and significantly variable can provide
constraints on the energy dependence of the variability, so we select
only sources for which the {\it rms\/} is $\ge 0.1$ at more than 3
sigma significance (determined from the uncertainty on the {\it rms},
which relates to the $\chi^2$ distribution as the variance is a sum of
squares, see e.g \citet{Done99}). This filtering leaves 19 AGNs.

We then excluded 2 known BL Lac objects (PKS2155-304 and 0716+714) since
their X-ray variablity is thought to be due to jet related processes.
For the two observations of NGC~4051, one (0157560101) only has an exposure
time of 49917 s, while the other (0109141401) does not have the pn${\_}$time
value in the catalog, so they were formerly excluded by the filtering.
However, since NGC~4051 is a well known, bright, NLS1 and our study is not
a statistical study, we still included these two observations. This
inclusion should not affect our general conclusions. We also included the
famous Seyfert 2 AGN, MCG-5-23-16 (0302850201), although its {\it rms\/}
variability is only 0.073. This gives us a total of 19 AGNs for subsequent study,
with one or more selected observations for each of them.

For each of these datasets we calculated the fractional variaibility
amplitude as a function of energy using the method of
\citet{Gierlinski06}. This showed the standard range of {\it rms\/}
spectral shapes. Only RX J0136.9-3510 (0303340101) displayed the very
different type of {\it rms\/} spectrum associated with the QPO AGN, RE
J1034+396, in which the fractional amplitude rises as a function of
energy, then remains high.  Having identified this unusual AGN, we
investigate its properties in more detail below.

\section{RX J0136-3510: lightcurve and {\it RMS\/} Spectrum}
\label{section-rms}

The lightcurve and {\it rms\/} spectrum are extracted from the X-ray
data (0.3-10 keV), using {\tt SAS7.1.0} and {\tt xronos5.21}. We
use regions with radius of 70{\arcsec}, 40{\arcsec} and 40{\arcsec}
for pn, MOS1 and MOS2, respectively.  No significant pile-up is seen
from the {\tt SAS} command {\tt epatplot} so no central region
was excluded. There are high background flares during the first
10000 s, so we exclude these data, resulting in $\sim$40000 s (with
background count rate$<0.4$ counts/s) for the {\it rms\/} spectrum
generation. Figure~\ref{lightcurve} shows the resultant total
lightcurve (PN, MOS1 and MOS2), binned on 200~s. The source shows
strong variability (also seen by Ghosh et al. 2009, in preparation),
with the {\it rms\/} fractional variation being
0.13. However, there is no obvious QPO, and a power spectral analysis shows no peaks above
even $1\sigma$ significance using the method of \citet{Vaughan05}.

We rebin the lightcurve on 2000~s in order to provide enough count statistics
to calculate the fractional variability as a function of energy.
Fig.~\ref{rms-spectrum} shows that this {\it rms} rises steeply from
0.3 keV to $\sim$2 keV, then flattens off.  The weak decrease from
2-10 keV is not significant as the low count rate means their are
large uncertainties on the last point.  We also calculated the {\it
rms\/} spectra from 200 s binning, but this is not significantly
different.

\section{Black Hole Mass Estimation}
\label{section-mass}

\begin{table*}
\begin{minipage}{140mm}
\caption{The best-fit parameters in our model for the optical to X-ray broad band SED.}
\label{parametertable}
\begin{tabular}{cccccccc}
\hline
 constant 2 & T$_{disk}$ (eV) & kT$_{e}$ (keV) & $\tau$ & N$_{comp}$ & ${\Gamma}_{pow}$ & N$_{pow}$ & $\chi^{2}_{\nu}$/d.o.f\\
\hline
 $0.92^{+0.10}_{-0.10}$ & $7.93^{+0.28}_{-0.31}$ & $0.28^{+0.03}_{-0.02}$ & $12.17^{+0.72}_{-0.75}$ & $10.64^{+2.37}_{-1.73}$ & $2.28^{+0.07}_{-0.08}$ & $1.12^{+0.55}_{-0.37}$ & 475/493\\
\hline
\end{tabular}
\end{minipage}
\end{table*}

The optical spectrum has an $Fe II/H\beta$
flux ratio of $\sim 8.3$(\citealt{Grupe99}), compared to the $\sim 1$ average for NLS1s
(\citealt{Veron01}), making RX J0136.9-3510 an unusual NLS1
\citep{Ghosh04}.We use the optical spectrum shown in
Fig~\ref{optical-SED} (D. Grupe, private
communication) to estimate the black hole mass. The
H$\beta$ line width is used as a proxy for this (see \citealt{Woo02} and references therein) from
\begin{equation}
\label{BH-mass}
M_{BH}=4.817{\times}[{{{\lambda}L_{\lambda}(5100{\AA})}\over{10^{44}erg s^{-1}}}]^{0.7}{FWHM}^{2}
\end{equation}
Comparing this method for a sample of AGN with reverberation mapping, the rms difference is about 0.5 dex
(\citealt{Woo02}). The flux at ${\lambda}=5100{\AA}$ from the optical spectrum is
${\sim}1.5{\times}10^{-16}$erg cm$^{-2}$s$^{-1}$. The luminosity distance
is  $D_{L}=1455.2$~Mpc for $z=0.289$ assuming
$H_{0}=72$ km Mpc$^{-1}$, ${\Omega}_{M}=0.27$, and ${\Omega}_{vac}=0.73$.

Estimating the FWHM of the line is not straightforward as the H$\beta$ line profile
is complex. There
is clearly also a component from the extended narrow line region (NLR). This
should be similar to the profile of the [OIII] emission line. So \citet{Grupe99}
use a template constructed from the [OIII]$\lambda$5007 line to represent the
narrow component, together with a broader Gaussian component, with
FWHM=1320 km/s, to reconstruct the $H\beta$ line.  This gives a black
hole mass estimate of $1.3{\times}10^{7}M_{\sun}$.

However, the optical spectrum plainly has limited
signal-to-noise, and the [OIII] 5007 line is probably contaminated by
FeII emission as it should be in 3:1 ratio with the (unobserved)
[OIII] line at 4959\AA. Hence we perform our own best fit of the H$\beta$ line
profile with a gaussian of width 870 km/s for the narrow line component, see Figure~\ref{optical-SED}.
This gives a FWHM for the broad component of 3200$\pm$2600 km s$^{-1}$,
with a (not significant) blueshift of -370$\pm$1100 km s$^{-1}$.
Using our new value for the FWHM, the resultant black hole mass is $7.85{\times}10^{7}M_{\sun}$.

Even with this lower black hole mass, this AGN is still probably $\sim 10\times$ more
massive than the QPO AGN, RE J1034+396, (with the previous mass estimate
being $\sim 50\times$ larger). Thus any similar QPO in RX J0136.9-3510 would be on
timescales 10-50$\times$ larger, requiring a much longer X-ray observation
in order to detect it.

\section{Broad band SED Analysis and Eddington Ratio}
\label{section-SED}

We use the standard products to obtain the XMM-Newton X-ray (PN)
spectra and optical/UV (OM) photometry.  We then combine these with
selected continuum points from the (non-simultaneous) optical spectum
(D. Grupe, private communication) using {\tt FLX2XSP} to incorporate
these into the same format as the XMM-Newton data. We likewise include
the J,H and K near infrared flux points from 2MASS, and perform all
spectral fitting using {\tt xspec11.3.2}.

We follow the approach of \citet{Vasudevan09} in modelling the
broadband SED, using {\tt diskpn} to model an accretion disc extending
down to the last stable orbit around a non-spinning black
hole. However, our source is at $z=0.289$ so we modify the {\tt
diskpn} code to incorporate the redshift dependance. The normalisation
of the {\tt diskpn} model is $(M^2\cos i) /(D_{kpc}^2\beta^4)$ where
$M$ is the mass in solar units, $D_{kpc}$ is the distance in kpc, and
the cosine of the inclination and colour temperature correction ($\cos
i$ and $\beta$, respectively) are both set to unity following
\citet{Vasudevan09}. We start at the highest estimated black hole mass
in Section~\ref{section-mass}, which gives a {\tt diskpn}
normalization of 2910, and fix this in the spectral fitting. Compton
scattering of these disc photons can be approximated by a broken power
law ({\tt bknpower}), with index of $\Gamma=0.33$ below a break at
$3kT_{disk}$. We then added a low temperature Comptonisation component
to model the soft excess, using {\tt comptt}, with seed photons set to
the disc temperature. We assume that these intrinsic components are
absorbed by both gas and dust in our Galaxy, and so fix this parameter
to the the Galactic HI
column\footnote{http://heasarc.gsfc.nasa.gov/cgi-bin/Tools/w3nh/w3nh.pl}
({\tt wabs}) value of $N_H=0.0208{\times}10^{22}$~cm$^{-2}$. The reddening
({\tt redden}) is linked to this assuming a E(B-V)=$1.74\times
10^{-22} N_H$ (\citealt{Spitzer78}). Since the optical data were not
simultaneous with the UV and X-ray data, we allow for long time-scale
variation as a constant offset in normalisation between the XMM-Newton
data and the optical spectrum. We exclude the infrared data points
from our spectral fitting, since the model we use describes the
intrinsic emission from the accretion flow whereas the infrared
emission is likely due to reprocessing of the UV emission by dust in
the host galaxy, plus a possible contribution from intrinsic
starlight.  The resultant best-fit parameters are given in
Table~\ref{parametertable}.

Figure~\ref{SED} shows the rebinned data (black), with the optical
flux corrected for their best fit normalisation of $\sim 0.92\times$
that of the XMM-Newton UV and X-ray spectra, and all datapoints
are corrected for absorption/reddening. This model is a good description
of the overall shape of the optical/UV/X-ray spectrum and gives a
bolometric (0.001-100 keV) flux of $1.1{\times}10^{-10}$erg
cm$^{-2}$s$^{-1}$, corresponding to a bolometric luminosity of
$2.7{\times}10^{46}$erg s$^{-1}$.

Using our own fitting value for the FWHM, from which the resultant
black hole mass is $7.85{\times}10^{7}M_{\sun}$, the Eddington ratio
for RX J0136.9-3510 is as high as $\sim 2.7$! Moreover, if we adopt
the lower estimated black hole mass which is
$1.3{\times}10^{7}M_{\sun}$, then the normalization of {\tt diskpn}
is 80, and the resultant disk temperature rises to 31~eV. The UV
region is then dominated by disk emission, though the
Comptonisation still is required to model
the soft X-ray excess. This gives a
bolometric flux of $8.8\times 10^{-11}$~ergs s$^{-1}$ cm$^{-2}$,
giving an even higher Eddington
ratio of 13.2! This is the highest known Eddington ratio for an AGN
(\citealt{Vasudevan09}; \citealt{Shen08}). Thus modelling the
spectral energy distribution with the two extreme mass estimates gives
a range for the Eddington ratio of RX J0136.9-3510 of 2.7-13.2. Even
without models, simply integrating the observed spectrum using a
straight line to connect the UV and soft X-ray data gives an Eddington
ratio of $\sim 1$ for the highest black hole mass, making this a
robustly super-Eddington source.

\section{Summary and Conclusion}

In this paper we report that RX J0136.9-3510 is the only well observed,
X-ray bright, variable AGN which has a similar energy dependence to its X-ray
variability as the so far unique QPO AGN, RE J1034+396.
Its Eddington ratio is similarly high, around 3, although its larger mass means any
QPO is undetectable in our data. Its broad band SED is also remarkably similar to that
of RE J1034+396, being well modelled by a low temperature, optically thick
Comptonisation of the accretion disc spectrum, plus a tail extending to higher energies.
Spectra such as this have also been fit by ``slim'' disc models (\citealt{Abramowicz89}),
where the accretion rate is so high that radiation cannot easily escape vertically
before it is carried radially (advected) along with the flow (\citealt{Puchnarewicz01};
\citealt{Wang03}). However, simple slim disc models do not fit the curvature of the
soft X-ray spectra as well as Comptonisation (\citealt{Middleton09}), although more
complex models of slim discs do include such scattering in the disc atmosphere
(e.g. \citealt{Kawaguchi03}).

Low temperature, optically thick Comptonisation is also occasionally seen in the stellar
mass black hole binary systems, for example in the most extreme mass accretion rate
spectra of GRS~1915+105
(\citealt{Middleton06}; \citealt{Middleton09}).
Recent studies of spectra of the Ultra-Luminous X-ray sources also indicates that
these are well modelled by such material (\citealt{Gladstone09}). This evidence suggests
that there is indeed a distinct spectral state which can only be attained by super
Eddington flows (\citealt{Gladstone09}). Future long duration X-ray observations of AGN
should reveal additional examples, and objects with
low black hole masses are potential QPO candiates.

\section*{Acknowledgements}
We acknowledge D. Grupe for permission to use the optical data.
C. C. Jin acknowledges financial support through Durham Doctoral Fellowship.


\begin{thebibliography}{}
\bibitem[\protect\citeauthoryear{Abramowicz, Kato \& Matsumoto}{1989}]{Abramowicz89}Abramowicz M. A., Kato S., Matsumoto R., 1989, PASJ, 41, 1215
\bibitem[\protect\citeauthoryear{Bian et al.}{2008}]{Bian08}Bian W., Hu C., Gu Q., Wang J., 2008, MNRAS, 390, 752
\bibitem[\protect\citeauthoryear{Boller, Brandt \& Fink}{1996}]{Boller96}Boller Th., Brandt W. N., Fink H., 1996, A{\&}A, 305, 53
\bibitem[\protect\citeauthoryear{Boroson}{2002}]{Boroson02}Boroson T. A., 2002, ApJ, 565, 78
\bibitem[\protect\citeauthoryear{Brandt, Mathur \& Elvis}{1997}]{Brandt97}Brandt W. N., Mathur S., Elvis M., 1997, MNRAS, 285, L25
\bibitem[\protect\citeauthoryear{Casebeer, Leighly \& Baron}{2006}]{Casebeer06}Casebeer D. A., Leighly K. M., Baron E., 2006, ApJ, 637, 157
\bibitem[\protect\citeauthoryear{Crummy et al.}{2006}]{Crummy06}Crummy J., Fabian A. C., Gallo L., Ross R. R., 2006, ApJ, 365, 1067
\bibitem[\protect\citeauthoryear{Done et al.}{1990}]{Done99}Done C., Ward M. J., Fabian A. C., Kunieda H., Tsuruta S., Lawrence A., Smith M.G., Wamsteker W., 1990, MNRAS, 243, 713
\bibitem[\protect\citeauthoryear{Edelson et al.}{1999}]{Edelson99}Edelson R., Vaughan S., Warwick R., Puchnarewicz E., George I., 1999, MNRAS, 307, 91
\bibitem[\protect\citeauthoryear{Fabian et al.}{2004}]{Fabian04}Fabian A. C., Miniutti G., Gallo L., Boller Th., Tanaka Y., Vaughan S., Ross R. R., 2004, MNRAS, 353, 1071
\bibitem[\protect\citeauthoryear{Gebhardt et al.}{2000}]{Gebhardt00}Gebhardt K. et al., 2000, ApJ, 539, L13
\bibitem[\protect\citeauthoryear{Gierli\'{n}ski \& Done}{2006}]{Gierlinski06}Gierli\'{n}ski M., Done C., 2006, MNRAS, 371, L16
\bibitem[\protect\citeauthoryear{Gierli\'{n}ski et al.}{2008}]{Gierlinski08}Gierli\'{n}ski M., Middleton M., Ward M., Done C., 2008, Nature, 455, 369
\bibitem[\protect\citeauthoryear{Ghosh et al.}{2004}]{Ghosh04}Ghosh K. K., Swartz D. A., Tennant A. F., Wu J., Ramsey B., 2004, ApJ, 607, L111
\bibitem[\protect\citeauthoryear{Gladstone, Roberts \& Done}{2009}]{Gladstone09}Gladstone J., Roberts T., Done C., MNRAS, submitted
\bibitem[\protect\citeauthoryear{Goodrich}{1989}]{Goodrich89}Goodrich R. W., 1989, ApJ, 342, 224
\bibitem[\protect\citeauthoryear{Grupe et al.}{1996}]{Grupe96}Grupe D. 1996, Ph.D. Thesis, Univ. G\"{o}ttingen
\bibitem[\protect\citeauthoryear{Grupe et al.}{1999}]{Grupe99}Grupe D., Beuermann K., Mannheim K., Thomas H. C., 1999, A{\&}A, 350, 805
\bibitem[\protect\citeauthoryear{Kawaguchi}{2003}]{Kawaguchi03}Kawaguchi T., 2003, ApJ, 593, 69
\bibitem[\protect\citeauthoryear{Kuraszkiewicz et al.}{2000}]{Kuraszkiewicz00}Kuraszkiewicz J., Wilkes B. J., Czerny B., Mathur S., 2000, ApJ, 542, 692
\bibitem[\protect\citeauthoryear{Larsson et al.}{2008}]{Larsson08}Larsson J., Miniutti G., Fabian A. C., Miller J. M., Reynolds C. S., Ponti G., 2008, MNRAS, 384, 1316
\bibitem[\protect\citeauthoryear{Leighly}{1999a}]{Leighly99a}Leighly K. M., 1999, ApJS, 125, 297
\bibitem[\protect\citeauthoryear{Leighly}{1999b}]{Leighly99b}Leighly K. M., 1999, ApJS, 125, 317
\bibitem[\protect\citeauthoryear{Magorrian et al.}{1998}]{Magorrian98}Magorrian J. et al., 1998, ApJ, 115, 2285
\bibitem[\protect\citeauthoryear{Mathur}{2000}]{Mathur00}Mathur S., 2000, MNRAS, 314, L17
\bibitem[\protect\citeauthoryear{Middleton et al.}{2006}]{Middleton06}Middleton M., Done C., Gierli\'{n}ski M., Davis S. W., 2006, MNRAS, 373, 1004
\bibitem[\protect\citeauthoryear{Middleton, Done \& Gierli\'{n}ski}{2007}]{Middleton07}Middleton M., Done C., Gierli\'{n}ski M., 2007, MNRAS, 381, 1426
\bibitem[\protect\citeauthoryear{Middleton et al.}{2009}]{Middleton09}Middleton M., Done C., Ward M., Gierli\'{n}ski M., Schurch N., 2009, MNRAS submitted (arXiv:0807.4847v1)
\bibitem[\protect\citeauthoryear{Morgan, Remillard \& Greiner}{1997}]{Morgan97}Morgan E. H., Remillard R. A., Greiner J., 1997, ApJ, 482, 993
\bibitem[\protect\citeauthoryear{Mullaney \& Ward}{2008}]{Mullaney08}Mullaney J. R., Ward M. J., 2008, MNRAS, 385,53
\bibitem[\protect\citeauthoryear{Osterbrock \& Pogge}{1985}]{Osterbrock85}Osterbrock D. E., Pogge R. W., 1985, ApJ, 297, 166
\bibitem[\protect\citeauthoryear{Petrucci et al.}{2007}]{Petrucci07}Petrucci P. O. et al., 2007, A{\&}A, 470, 889
\bibitem[\protect\citeauthoryear{Ponti et al.}{2006}]{Ponti06}Ponti G., Miniutti G., Cappi M., Maraschi L., Fabian A. C., Iwasawa K., 2006, MNRAS, 368, 903
\bibitem[\protect\citeauthoryear{Pounds, Done \& Osborne}{1995}]{Pounds95}Pounds K. A., Done C., Osborne J. P., 1995, MNRAS, 277, L5
\bibitem[\protect\citeauthoryear{Puchnarewicz et al.}{1992}]{Puchnarewicz92}Puchnarewicz E. M. et al., 1992, MNRAS, 256, 589
\bibitem[\protect\citeauthoryear{Puchnarewicz et al.}{1995}]{Puchnarewicz95}Puchnarewicz E. M., Mason K. O., Siemiginowska A., Pounds, K. A., 1995, MNRAS, 276, 20
\bibitem[\protect\citeauthoryear{Puchnarewicz et al.}{2001}]{Puchnarewicz01}Puchnarewicz E. M., Mason K. O., Siemiginowska A., Fruscione A., Comastri A., Fiore F., Cagnoni I., 2001, ApJ, 550, 644
\bibitem[\protect\citeauthoryear{Remillard \& McClintock}{2006}]{Remillard06}Remillard R. A., McClintock J. E., 2006, ARAA, 44, 49
\bibitem[\protect\citeauthoryear{Shen et al.}{2008}]{Shen08}Shen Y., Greene J. E., Strauss M. A., Richards G. T., Schneider D. P., 2008, ApJ, 680, 169
\bibitem[\protect\citeauthoryear{Spitzer}{1978}]{Spitzer78}Spitzer L. Jr., 1978, JRASC, 72, 349
\bibitem[\protect\citeauthoryear{Vasudevan \& Fabian}{2009}]{Vasudevan09}Vasudevan R. V., Fabian A. C., 2009, MNRAS, 392, 1124
\bibitem[\protect\citeauthoryear{Vaughan et al.}{2003}]{Vaughan03}Vaughan S., Edelson R., Warwick R. S., Uttley P., 2003, MNRAS, 345, 1271
\bibitem[\protect\citeauthoryear{Vaughan et al.}{2004}]{Vaughan04}Vaughan S., Fabian A. C., 2004, MNRAS, 348, 1415
\bibitem[\protect\citeauthoryear{Vaughan}{2005}]{Vaughan05}Vaughan S., 2005, A{\&}A, 431, 391
\bibitem[\protect\citeauthoryear{V\'{e}ron-Cetty, V\'{e}ron \& Goncalves}{2001}]{Veron01}V\'{e}ron-Cetty M., P., V\'{e}ron P., Goncalves A. C., 2001, A{\&}A, 372, 730
\bibitem[\protect\citeauthoryear{Wang \& Netzer}{2003}]{Wang03}Wang J.-M., Netzer H., 2003, A\&A, 398, 927
\bibitem[\protect\citeauthoryear{Wills et al.}{1999}]{Wills99}Wills B. J., Laor, A., Brotherton M. S., Wills, D., Wilkes, B. J., Ferland, G. J., Shang Z., 1999, ApJ, 515, L53
\bibitem[\protect\citeauthoryear{Winter et al.}{2009}]{Winter09}Winter L. M., Mushotzky R. F., Reynolds C. S., Tueller J., 2009, ApJ, 690, 1322
\bibitem[\protect\citeauthoryear{Woo \& Urry}{2002}]{Woo02}Woo Jong-hak, Urry C. M., 2002, ApJ, 579, 530
\end{thebibliography}
\end{document}